\documentclass[11pt]{amsart}

\usepackage{amsmath,amsfonts}
\usepackage{latexsym,amssymb}
\usepackage{bm}

\newcommand{\dd}{\mathtt{d}}

\begin{document}

\title[On incorrectness of application of the Helmholtz decomposition...]
{On incorrectness of application of the Helmholtz decomposition to microscopic electrodynamics}

\author{Vladimir~Onoochin}
\address{Sirius, 3A Nikoloyamski lane, Moscow, Russia.}

\begin{abstract}
The integral expressions served to decompose vector field into irrotational  and divergence-free components represent modern version of the Helmholtz decomposition theorem. These expressions are also widely used to decompose the electromagnetic fields.  

However, an appropriate analysis of application of these expressions to electrodynamics shows that the improper integral arising in the procedure for calculating these components makes such a decomposition impossible.
\end{abstract}

\maketitle
   
\section{Introduction}

The  Helmholtz  decomposition theorem states that any vector function ${\bf F}({\bf r})$, defined in unbounded region and vanishing suitably quickly at infinity ($r\to\infty$),
can be expressed as a sum of an irrotational (curl-free) vector and a divergence-free (divergence-free) vector in the form 
\begin{equation}
{\bf F}({\bf r})=\bm{\nabla} \Phi ({\bf r})+ [\bm{\nabla}\times{\bf C}({\bf r})] \,.
\end{equation}
where $ \Phi ({\bf r})$ and ${\bf C}({\bf r})$ are some function (that not always represent the potentials) satisfying the conditions $\bm{\nabla}\cdot {\bf F}=\Phi$, $\bm{\nabla}\times{\bf F}={\bf C}$.

The  theorem was introduced by Hermann Helmholtz  in a paper on vortex motion (in a hydrodynamic context) in 1858. Due to its generality, the theorem finds application in various areas of physics, from hydrodynamics to electromagnetism. Because of the properties of the electromagnetic field (the magnetic field is purely divergence-free and the electric field can be split into bound component of irrotational type and the radiated component of divergence-free type) it was a 'natural way' to apply the Helmholtz theorem to decompose the EM fields, first of all for practical applications. Since the radiated EM field component is of the divergence-free type, the separation of the radiated component, which is more important for applications, from the coupled or potential component allows us to analyze the behavior of the radiated fields even in the near zone of the antenna.

But the application of the decomposition procedure to electrodynamics has encountered certain difficulties. Formally, this theorem is not appropriate for the decomposition of electromagnetic fields, since the radiated components of these fields do not decrease sufficiently quickly at infinity but drops as $1/r$. However, there are a number of works~\cite{Petr} where it is shown that, under certain conditions, the decomposition procedure can also be performed for radiated fields.

Meanwhile, the main obstacle in applying the theorem to the decomposition of an electric field is the presence of divergences at the points where the sources are located. It should be noted that no one has studied this aspect of the theorem before.

\section{Analysis of the Helmholtz decomposition}

It seems obvious that any vector field can be represented as the sum of its two components; one of them is of irrotational type and the other is of divergence-free type. But if so, two questions arise\newline
- how to find these components?\newline
- how to prove that the procedure for obtaining these components gives an unambiguous result?

The Helmholtz decomposition gives the answer to both questions. It should be noted that the Helmholtz theorem is represented by Eq.~(1). But the expressions that allow to find the quantities $ \Phi ({\bf r})$ and ${\bf C}$ form so called {\it Helmholtz decomposition} ~\cite{Bh}, or the inverse problem to the Helmholtz theorem.

Actually, this theorem is based on the vector identity
\begin{equation}
\bm{\nabla}^2{\bf F}=\bm{\nabla}(\bm{\nabla}\cdot{\bf F})-\left[\bm{\nabla}\times[\bm{\nabla}\times{\bf F}]\right]\,, 
\label{diff}
\end{equation}
and it is easy to see that the terms in the brackets are of irrotational and divergence-free type. 

It is known that for any regular function ${\bf F}({\bf r})$ the identity 
\[
{\bf F}({\bf r})=\frac{1}{4\pi}\int \frac{\nabla'^2{\bf F}({\bf r}')}{|{\bf r}-{\bf r}' | }\dd ^3r'\,,
\]
is fulfilled. By means of~(\ref{diff}) this identity can be 
 written as
 \begin{equation}
{\bf F}({\bf r})=-\frac{1}{4\pi}\int \frac{  \bm{\nabla}'\left(\bm{\nabla}'\cdot{\bf F}({\bf r}')\right)}
{|{\bf r}-{\bf r}' |} \dd ^3r'+
\frac{1}{4\pi} \int \frac{\left[ \bm{\nabla}'\times[\bm{\nabla}'\times{\bf F}({\bf r}')]\right]}{|{\bf r}-{\bf r}' |}\dd ^3r'
\,.\label{bsc}
\end{equation}
In~Eq.~(\ref{bsc}), 'primed' operators act on the internal variable $r'$. Despite, formally one is able to find decomposed components of the vector field ${ \bf F}$, presence of irrotational and divergence-free terms in numerators of the integrands but not in the integrands does not allow to conclude that the integrals represent the decomposed components. So one more step in procedure is needed to obtain representation of the theorem in well known form (Eqs. (5.7-7) and (5.7-8) of~\cite{KK})
\begin{equation}
{\bf F}({\bf r})=-\bm{\nabla}\frac{1}{4\pi}\int \frac{\bm{\nabla}'\cdot{\bf F}({\bf r}')}
{|{\bf r}-{\bf r}' |}\dd ^3r'+
\bm{\nabla}\times \frac{1}{4\pi}\int \frac{\bm{\nabla}'\times{\bf F}({\bf r}')}{|{\bf r}-{\bf r}' |}\dd ^3r'
\,.\label{HD}
\end{equation}
where non-'primed' operators act on the external variable $r$.
But this form of the theorem has one weak point, namely, the integrals diverge at infinity if ${\bf F}\simeq 1/r$, which corresponds to the radiated fields in classical electrodynamics. Therefore, the form  
\begin{equation}
{\bf F}({\bf r})=-\frac{1}{4\pi}\int \bm{\nabla}\left(\frac{\bm{\nabla}'\cdot{\bf F}({\bf r}')}
{|{\bf r}-{\bf r}' |}\right) \dd ^3r'+
\frac{1}{4\pi} \int \bm{\nabla}\times\frac{\bm{\nabla}'\times{\bf F}({\bf r}')}{|{\bf r}-{\bf r}' |}\dd ^3r'
\,.\label{basic}
\end{equation}
is more appropriate since it provides finite value of the integrals~\cite{Petr}, that is actual for calculating the radiated EM fields. 

Correctness of Eq.~\eqref{basic} is proven by transformation of the integrals
\begin{equation}
-\frac{1}{4\pi}\int \frac{\bm{\nabla}'\left(\bm{\nabla}'\cdot{\bf F}({\bf r}')\right)}{|{\bf r}-{\bf r}' |}\dd ^3r'
 =\frac{1}{4\pi}\int \bm{\nabla}\left(\frac{\bm{\nabla}'\cdot{\bf F}({\bf r}')}
 {|{\bf r}-{\bf r}' |}\right) \dd ^3r\,, \label{o1}
\end{equation}
\begin{equation}
 \frac{1}{4\pi}\int \bm{\nabla}\times\frac{\bm{\nabla}'\times{\bf F}({\bf r}')}{|{\bf r}-{\bf r}' |}\dd ^3r'=
\frac{1}{4\pi}\int  \left [\bm{\nabla}\times\left[\bm{\nabla}\times \frac{{\bf F}({\bf r}')}
{|{\bf r}-{\bf r}' |} \right]\right]\dd ^3r' \,, \label{o2}
\end{equation}
These relations are proven using integration by parts. Details of these integrations for vector functions are explained in~\cite{Woo}, Eqs. (9) to (13). In this proof, one should be sure that the relations~(\ref{o1}) and (\ref{o2}) are mathematically correct. 

For the hydrodynamics, it is correct, but not for the electrodynamics. In microscopic electrodynamics, the EM fields are created by classical charges $q$ and their spatial dependence is of the form
\begin{equation}
{\bf E}({\bf r})\simeq q\left(\frac{{\bf a}(t)}{r^2}+ \frac{{\bf b}(t)}{r}\right)\,, \label{form}
\end{equation}
(${\bf a}$ and ${\bf b}$ are some vector quantities which can depend on the velocity and acceleration of the charge) where the first term in~(\ref{form})  represents the bound (near-zone) field and the second term -- the radiated field.

However, if one inserts the expression for electric field in the form~(\ref{form}) into the {\it lhs} of Eq.~(\ref{o1}) one obtains the divergent term in the integrand since $\bm{\nabla}\cdot {\bf E}\simeq 1/r^3$ and therefore, $\bm{\nabla}(\bm{\nabla}\cdot {\bf E})\simeq 1/r^4$.

Here, it is reasonable to clarify one aspect of application of the Helmholtz theorem. Its tool was developed in the XIX century when the classical electrons are not treated as point-like particles. The charge was considered as an object distributed in space. Therefore it was accepted as an obvious fact that calculation of the EM fields and even of their derivatives does not yield singularities.

However, it is not so. Convergence of integrals containing derivatives of the Coulomb-like potentias (and derivatives of the Coulomb-like fields) created by distributed sources is considered by Tikhonov (Ch. IV-5.5 'Second derivatives of the volume potential'~\cite{Tik}). 

Such integrals calculated over the volume $T$ and containing the derivatives of potential  
\[ 
\iiint_T\varrho(P)\frac{\partial^2}{\partial x^2}\frac{1}{R_{MP}} \dd \tau \,,
\]
where $\varrho(P)$ is the density of the source at a point $P$, and $R_{MP}$, the distance between $P$ and the point $M$ of calculating the integral, are of improper type. 

Despite the integral  does not absolutely converge, it conditionally converges. However, the integrals containing the third derivative of potential diverges even for distributed source. 

Therefore, transformation of the {\it lhs} to the {\it rhs} of Eq.~(\ref{o1}) cannot be made correctly. As a result, one cannot reduce {\it rhs} of Eq.~(\ref{basic}) to the integrals~(\ref{o1}) and~(\ref{o2}) written for the electric field, or
 \begin{eqnarray}
 -  \frac{1}{4\pi}\int \frac{\bm{\nabla}'\left(\bm{\nabla}'\cdot{\bf E} ({\bf r}')\right)}
{|{\bf r}-{\bf r}' |}\dd ^3r'+
\frac{1}{4\pi}\int \frac{\left[\bm{\nabla}'\times[\bm{\nabla}'\times{\bf E} ({\bf r}')]\right]}
{|{\bf r}-{\bf r}' |}\dd ^3r' \neq  \nonumber\\
 \neq \frac{1}{4\pi}\int \bm{\nabla}\left(\bm{\nabla}'\cdot \frac{{\bf E}({\bf r}') }{|{\bf r}-{\bf r}' |} \right)\dd ^3 r'  -
\frac{1}{4\pi}\int  \left [\bm{\nabla}\times\left[\bm{\nabla}'\times \frac{{\bf E}({\bf r}')}{|{\bf r}-{\bf r}' |} \right]\right]\dd ^3r' =\nonumber\\
= \frac{1}{4\pi} \int   {\bf E}({\bf r}') \,\nabla^2\frac{1} {|{\bf r}-{\bf r}' |}
\dd ^3r'= \int {\bf E}({\bf r}') \delta^{(3)}\left( {\bf r}-{\bf r}' \right) \dd^3 r'\,.
\end{eqnarray}
Therefore, Eq.~\eqref{basic} is not fulfilled for the electric fields and its decomposition onto irratotional and divergence-free components is questionable.

One can see that the obstacle that prevents to decompose the electric field by means of the integrals~(\ref{basic}) arises from the fact that the electric field has stronger singularity than the term $ 1/|{\bf r}-{\bf r}'|$.

Meanwhile, it follows from Tikhonov's analysis that in expression for the Helmholtz decomposition in a form~(\ref{bsc}) does not contain divergences and it looks like this form of decomposition -- under certain assumptions -- could be used for the EM fields. But the integral 
\[
\int \bm{\nabla}\left(\frac{\bm{\nabla}'\cdot{\bf E}({\bf r}')}
 {|{\bf r}-{\bf r}' |}\right) \dd ^3r\,\,\to\,\,
\int  \bm{\nabla}\left(\frac{\rho}{|{\bf r}-{\bf r}' |}\right)\dd ^3r'\
\]
is still of improper type. Therefore, operation of integration by parts is not well-defined for it. This means that chain of transformations of the electric fields is also mathematically incorrect,
 \begin{eqnarray}
 -\bm{\nabla}  \frac{1}{4\pi}\int \frac{\bm{\nabla}'\cdot{\bf E} ({\bf r}')}{|{\bf r}-{\bf r}' |}\dd ^3r'+
\bm{\nabla}\times  \frac{1}{4\pi}\int \frac{\bm{\nabla}'\times{\bf E} ({\bf r}')}{|{\bf r}-{\bf r}' |}\dd ^3r'
=\nonumber\\
 =-\bm{\nabla}\int \frac{\rho({\bf r}')} {|{\bf r}-{\bf r}' |}\dd ^3r' + \frac{1}{4\pi}\int 
 \left [\bm{\nabla}\times\left[\bm{\nabla}\times \frac{{\bf F}({\bf r}')}{|{\bf r}-{\bf r}' |} \right]\right]\dd ^3r' \neq \nonumber\\
 \neq \frac{1}{4\pi}\int \bm{\nabla}\left(\bm{\nabla}\cdot \frac{{\bf E}({\bf r}') }{|{\bf r}-{\bf r}' |} \right)\dd ^3 r'  -
\frac{1}{4\pi}\int  \left [\bm{\nabla}\times\left[\bm{\nabla}\times \frac{{\bf E}({\bf r}')}{|{\bf r}-{\bf r}' |} \right]\right]\dd ^3r' =\nonumber\\
=\frac{1}{4\pi} \int   {\bf E}({\bf r}') \,\nabla^2\frac{1} {|{\bf r}-{\bf r}' |}
\dd ^3r'= \int {\bf E}({\bf r}') \delta^{(3)}\left( {\bf r}-{\bf r}' \right) \dd^3 r'\,.
\end{eqnarray}
 
\section{Physical limitation on decomposition of the EM fields in microscopic electrodynamics}

The direct application of expressions for calculating the irrotational or divergence-free components (Eqs.~(\ref{bsc}) and ~(\ref{basic})) in classical electrodynamics is of little importance. The reason for this is that exact expressions are known only for two EM fields, of uniformly moving and uniformly accelerated charges. The decomposition of these fields is not as productive. More valuable would be the expressions allowing to calculate the irrotational and divergence-free components using sources of EM fields, without solving Maxwell's equations.

Such expressions are to describe the following processes:\newline
the scalar sources create the irrotational components of the vector fields, the vortex sources create the divergence-free components. In the case where these sources act independently of each other, for example, in electrostatic and magnetostatic, the Helmholtz theorem is successfully applied. 

In electrostatic, the sources of the electric field are free charges being at rest. In magnetostatic, the sources are the closed current loops for which div ${\bf J}=0$ and curl ${\bf J}\neq 0$. Divergence-free source is able to created only divergence-free static fields.

But the situation is quite different for moving charges which velocity and direction of motion vary with time.  Let us consider how radiated EM waves, which are of divergence-free type, are created. As it is shown by Jefimenko~\cite{Jef}, expression describing these waves is 
(Eq.~(2-2.12) of~\cite{Jef})
\begin{equation}
{\bf E}_{rad}=\frac{1}{c^2}\int \left\{\frac{\mathbf{r}-\mathbf{r}'}{|\mathbf{r}-\mathbf{r}'|^3}
\left((\mathbf{r}-\mathbf{r}')\cdot\left[\frac{\partial \mathbf{j}}{\partial t}\right]_{ret}
\right)-  \frac{1}{|\mathbf{r}-\mathbf{r}'|}\left[\frac{\partial \mathbf{j}}{\partial t}\right]_{ret}\right\}\dd ^3r'\,\,,\label{J}
\end{equation}
where subscription $\{ret\}$  means that the quantities depend on the retarded time.

One can see that the second term in the {\it rhs} of~(\ref{J}) is due to change of the current density ${\bf j}$ with time. But the first term in the above equation arises from the integral
\[
\int \frac{[\bm{\nabla}\rho]_{ret}}{|\mathbf{r}-\mathbf{r}'|}\dd^3r' \,.
\]
Hence not the only but two sources are responsible for creation of the divergence-free component of the EM field.
Physically it means that both sources of the EM fields, $\rho$ and ${\bf j}$ when changing in space and time, create spherical waves (for example, the above expression describes the spherical waves). Superposition of these spherical waves is such that they mutually compensate each the other in the longitudinal to (instant) direction of the charge motion. It is in certain disagreement with decomposition by means of the Helmholtz theorem -- in all works on application of this theorem, the only source is used to create a component of one type.

Typical approach to decomposition of the vector fields ${\bf F}$ is that only source of divergence-free type ($\bm{\nabla}\cdot{\bf C}=0$) is able to create the divergence-free component of ${\bf F}$. Similar approach is used by Jackson in description of the Coulomb gauge properties. The transverse current in the form~(Eq. (6.28) of~\cite{JDJ}))
\[
{\bf J}_t=\bm{\nabla}\times\bm{\nabla}\times\int \frac{{\bf J}({\bf r}' )}{|\mathbf{r}-\mathbf{r}'|}\dd^3r' \,\,;
\quad \bm{\nabla}\cdot{\bf J}_t=0\,.
\]
creates the divergence-free vector potential ${\bf A}_C$ in the Coulomb gauge since in this gauge $\bm{\nabla}\cdot{\bf A}_C=0$. 

It should be noted that Jefimenko's expression for the electric field is more general since it is derived directly for the EM fields, without using potentials. In other words, Jefimenko's expression does not depend on any gauge. Meanwhile it is difficult, if possible, to extract a source of divergence-free type from a sum of $[\nabla\rho]_{ret}$ and $[\partial_t{\bf j}]_{ret}$. Without extracting such a source from these quantities, the application of the Helmholtz decomposition to EM fields would be questionable. According to the author's knowledge, there is no one paper on this subject where this procedure would be done.

As a demonstration that the transformations~(\ref{o1}) and~(\ref{o2}) do not work for the electric field, let us show that application of 'weakened form' of the Helmholtz decomposition, Eq.~(\ref{basic}), to the EM fields leads to unphysical result. To do it, let us analyze the second integral in the {\it r.h.s} of Eq.~(\ref{basic}). 

Since the term $\bm{\nabla}'\times{\bf E} ({\bf r}')$ in the nominator of this integral describes only radiated component, the integral does not contain divergences. By means of the Maxwell equation $\bm{\nabla}\times{\bf E}=-(1/c)\partial_t{\bf B}$, this integral is transformed to
\begin{equation}
 \int \bm{\nabla}\times \frac{\bm{\nabla}'\times{\bf E} ({\bf r}')}{4\pi|{\bf r}-{\bf r}' |}\dd ^3r'=-
\int \bm{\nabla}\times  \frac{\partial_t{\bf B}({\bf r}')}
{4\pi c|{\bf r}-{\bf r}' |}\dd ^3r' = -\int \frac{\partial }{\partial t}\frac{\bm{\nabla}'\times{\bf B} ({\bf r}')}
{4\pi c|{\bf r}-{\bf r}' |}\dd ^3r'\,.
\end{equation}
The last integral in the above expression is obtained by introducing the operator $\nabla$ into the integral and removing the operator $\dfrac{\partial}{\partial t}$ out of the sign of the integral (in order to keep its finite value of the integral). The consideration of procedure of integration by part in introducing the operator $\bm{\nabla}\,\to\bm{\nabla}'$ to the nominator of the integral is omitted.

The last step is to apply the Maxwell equation $\bm{\nabla}\times{\bf B}=(1/c)\partial_t{\bf E}+(4\pi/c){\bf j}$ to the third integral. So finally Eq.~(\ref{basic}) written for the electric field takes a form 
 \begin{eqnarray}
 {\bf E}=-\int \bm{\nabla}\left(\frac{\bm{\nabla}'\cdot{\bf E} ({\bf r}')}{4\pi|{\bf r}-{\bf r}' |}\right)\dd ^3r'+
 \int \bm{\nabla}\times\frac{\bm{\nabla}'\times{\bf E} ({\bf r}')}{4\pi|{\bf r}-{\bf r}' |}\dd ^3r'=\nonumber\\
={\bf E}_{Coul}+\frac{1}{c^2}\int \frac{\partial^2}{\partial t^2}\frac{{\bf E}({\bf r}')} {4\pi|{\bf r}-{\bf r}' |}\dd ^3r' +
\frac{1}{c}\int \frac{\partial {\bf j}/\partial t} {|{\bf r}-{\bf r}' |}\dd ^3r'\,,\label{fin}
\end{eqnarray}
where ${\bf E}_{Coul}$ is the instantaneous Coulomb field. 

The second term in the {\it rhs} of Eq.~(\ref{fin}) is written in such a form in order to avoid divergences in integrals. But the main problem is due to appearance of the third term. It drops at infinity as $1/r$ so it should describe the radiated field (due to the time derivative of the current). However, it should describe the {\it longitudinal $E$ wave} because direction of the component given by this term coincides with the direction of the time derivative of the current.

Similar  expression (without the integral containing the second time derivative of the electric field in the nominator of ) was given by Jefimenko (Eq. (2-2.1) of~\cite{Jef}),
\begin{equation}
{\bf E}=\int \frac{\left[\bm{\nabla}\rho \right]_{ret}} {|{\bf r}-{\bf r}' |}\dd ^3r' +
\frac{1}{c}\int \frac{\left[\partial {\bf j}/\partial t\right]_{ret}} {|{\bf r}-{\bf r}' |}\dd ^3r'\,,\label{Jef}	
\end{equation}	
But the essential difference between Efimenko's expression and the expression~(\ref{fin}) is that the retarded time dependence of the sources in~(\ref{Jef}) leads to elimination of the longitudinal far-field component from the total electric field. There is no factor that can remove this component from a (hypothetical) electric field described by~(\ref{fin}). Therefore, the presence of a unphysical term should be treated as incorrectness of decomposition of the electric field if one applies the Helmholtz theorem in the form~(\ref{basic}) to the electrodynamics.

\section{Conclusion}

It is shown in this work that the EM field created by classical charges whose velocities and directions of motion vary with time (non-static case) cannot be decomposed into irrotational and devergence-free components. So the formulation of the Helmholtz decomposition theorem, in its modern version, needs at least in certain correction. In a case the corresponding expressions will be derived, it is desirable to verify their correctness by a closed form calculations. 

One point should be clarified here, namely, why this problem does not appear in hydrodynamics.

Despite the hydrodynamical systems also contain sources, the influence of these sources can be replaced by the boundary conditions specified given for them. In opposite,  the boundary of the classical charge cannot be strictly defined. As it is noted in Ch. 19-1 of~\cite{PP}, "Now in classical electrodynamics the only thing known about the electron is that it has a certain total charge, and any calculation of its radiation field cannot involve details of how this charge may be distributed geometrically within the electron". Therefore, the classical charges cannot be removed from consideration of the electrodynamic system by change them to the boundary conditions.

\end{document}